\documentclass[a4paper]{article}

\usepackage[english]{babel}
\usepackage[utf8x]{inputenc}
\usepackage[T1]{fontenc}

\usepackage[a4paper,top=3cm,bottom=2cm,left=3cm,right=3cm,marginparwidth=1.75cm]{geometry}

\usepackage{amsmath}
\usepackage{amssymb}
\usepackage{amsfonts}
\usepackage{graphicx}
\usepackage[colorinlistoftodos]{todonotes}
\usepackage[colorlinks=true, allcolors=blue]{hyperref}

\title{On Time-frequency Scattering and Computer Music}
\author{Vincent Lostanlen, New York University}
\date{First version: October 2018. Latest revision: May 2019.}

\begin{document}
\maketitle

\begin{quote}
\emph{... qu’il disperse le son dans une pluie aride ...}

\qquad \qquad --- St\'ephane Mallarm\'e
\end{quote}

The quest for an adequate representation of auditory textures lies at the foundation of computer music research.
Indeed, none of its analog predecessors ever managed a practical compromise between two concurrent needs in sound design:
first, to faithfully reproduce any pre-existing texture;
and secondly, to offer enough flexibility for sculpting novel textures from scratch.
For example, Schaeffer's \emph{musique concr\`{e}te} offered a precise typology of musical objects, yet constrains the composer to a figurativistic raw material \cite{schaeffer2017book}.
On the other hand, Stockhausen's \emph{Elektronische Musik}, as it arranges simple noises and tones through time, may have uncovered new avenues in musical abstraction; yet at the cost of a narrow, distinctively ``robotic'' timbral palette \cite{harvey1975book}.
In the history of music technology, such an opposition between specificity and expressivity is reflected in the respective developments of granular synthesis and additive synthesis: one is universal but computationally intractable, the other is terse but somewhat clunky.
With the democratization of analog-to-digital audio conversion, both aforementioned schools of thought came to decline, and new tools for sound manipulation in the time-frequency domain, such as the phase vocoder, gained momentum among contemporary music composers.
However, the progressive digitization of the music studio brought little progress to the long-lasting problem of audio texture synthesis and manipulation.

The science of auditory neurophysiology paved the way towards a computational framework for audio texture modeling that could reconcile the specificity of \emph{musique concr\`{e}te} with the expressivity of \emph{Elektronische musik}.
In 1996, Nina Kowalski and her colleagues employed an array of silicon electrodes to measure the cortical responses of a ferret to computer-generated ripple stimuli, exhibiting modulations both in time and frequency \cite{kowalski1996jneuro}.
Pairwise correlations between stimuli and responses led to an exhaustive mapping of the primary auditory cortex of mammals, which associates each neuron to a spectrotemporal receptive field (STRF) — that is, the time-frequency representation pattern eliciting maximal excitation of this neuron.
What Kowalski et al.~concluded is that our brain integrates the acoustic spectrum through time in terms of its spectrotemporal modulations at various scales (pitch intervals) and rates (pulse tempi).
Neither exclusively rhythmic (temporal), nor exclusively harmonic (frequential), our brain is indeed a joint, rhythmico-harmonico-melodic processor that encodes sound into a multifaceted sensation.

Despite marking a watershed in our understanding of music perception, this finding long remained outside the technological landscape of computer music designers, because the biologically inspired STRF representation was not an invertible procedure.
Instead, although STRF allowed to map sounds to specific areas of the auditory cortex, the dual problem of sonifying the neuroelectrical activations of these areas had remained largely unexplored.
In addition, since STRF had been obtained empirically from ferret neuronal action potentials, the resulting representation could not be interpreted \emph{post hoc} in terms of continuous perceptual parameters, such as pitch or tempo.
Simply put, STRF are more concrete than \emph{musique concr\`{e}te} itself — in lieu of eardrum vibrations, what they contain is a heatmap of primary auditory cortex activity — but lack the mathematical concision of an \emph{Elektronische Musik} score in order to allow for any compositional intervention on the world of natural sounds.

From 2013 to 2016, I was a grad student at \'{E}cole normale sup\'{e}rieure, striving to develop new convolutional operators in the time-frequency domain for modeling musical timbre \cite{lostanlen2017phd}.
With my coworker Joakim And\'{e}n and my advisor St\'{e}phane Mallat, I contributed to a STRF-based computational model for audio texture synthesis, under the name of time-frequency scattering.
Time-frequency scattering was meant as the successor to ``time scattering'', as it was formulated by Mallat himself in 2012.
The name was coined as a nod to the world of quantum mechanics: from the reddish shade of a sunset to the glistening of a pearl, the umbrella term of scattering encompasses many different microscopic phenomena.
The commonality between these phenomena is that they all involve a radiation of some kind as well as a maze of nonuniformities.
Let $\boldsymbol{g}$ be a Gaussian bell curve. In the context of scattering transforms, the radiation is a sound pressure wave $\mathbf{U}_0(t)$ while the maze consists of Morlet wavelets 
\begin{equation}
\boldsymbol{\psi}_\gamma (t) =
2^\gamma \boldsymbol{g}(2^\gamma t) \big[ \exp(2\pi \mathrm{i} 2^\gamma t)  - \boldsymbol{\hat{g}}(2^\gamma) \big]
\label{eq:psi}
\end{equation}
tuned at resolutions $2^\gamma$, as well as modulus nonlinearities.

Before time-frequency scattering was formalized, Mallat had defined the time scattering transform as a cascade of purely temporal wavelet modulus operators:
\begin{equation}
\mathbf{U}_{m+1}(t, \gamma_1 \ldots \gamma_{m+1}) =
\big\vert \mathbf{U}_m \overset{t}{\ast} \boldsymbol{\psi}_{\gamma_{m+1}} \big\vert (t)
=
\left\vert
\int_{-\infty}^{+\infty} \mathbf{U}_m(\tau,\gamma_1 \ldots \gamma_m) \boldsymbol{\psi}_{\gamma}(t-\tau) \;\mathrm{d}\tau
\right\vert
\end{equation}
and then generalized his theory to all real-valued functions of finite energy defined over the irreducible representations of a given compact Lie group \cite{mallat2012cpam}.
Shortly thereafter, my coworker Ir\`{e}ne Waldspurger proved that scattering transforms, despite the loss of phase incurred by the complex moduli, are invertible with continuous inverse \cite{waldspurger2015phd}.
She resorted to advanced methods in topology and complex analysis (namely the Riesz-Fr\'{e}chet-Kolmogorov theorem and meromorphic extensions, among others) to come up with this astonishing result:
on the condition that the chosen wavelets form a ``tight'' frame of the functional space at hand, and towards the limit of infinite depth $m\rightarrow\infty$, the time variable can ultimately be removed from the equation, because the oscillatory nature of sound vibrations in $\mathbf{U}_0 (t)$ gets fully characterized by its interference pattern through the scattering network.
Going back to the metaphor of Mie scattering in quantum mechanics, it is as though Mallat and Waldspurger had unearthed some kind of all-witnessing crystal, whose eternal glisten were a petrified testimony of every light it had seen before.

Waldspurger's invertibility theorem spurred my interest for improving the state of the art in audio texture synthesis.
Nevertheless, one important drawback of the scattering transform --- in its original, purely temporal definition --- is that it does not include the notions of relativity of pitch nor relativity of tempo.
Instead, each wavelet modulus layer decomposes all paths $p = (\gamma_1 \ldots \gamma_m)$ asynchronously.
It was after personal communications with Shihab Shamma that we realized the crucial importance of accounting for joint modulations in time and frequency ($t$ and $\lambda_1 = 2^{\gamma_1}$); or, said in algebraic terms, for elastic displacements over the affine Weyl-Heisenberg group on $\mathbf{L}^2(\mathbb{R})$.
Consequently, we proceeded to generalize the one-dimensional Morlet wavelet in Equation \ref{eq:psi} by a tensor product over multiple variables $(v_1 \ldots v_R)$, yielding time-frequency scattering wavelets of the form
\newcommand{\CONS}{\!::\!}
\begin{equation}
\boldsymbol{\Psi}_{\lambda} (v_1\ldots v_R) =
\bigotimes_{r=1}^{R}
2^{\gamma \,\CONS\, v_r}
(\theta \CONS v_r)
\boldsymbol{g}_r(2^{\gamma \,\CONS\, v_r} v_r)
\Big[
\exp\big(2\pi\mathrm{i}2^{\gamma \,\CONS\, v_r} (\theta \CONS v_r) v_r\big)
- \boldsymbol{\hat{g}}_r\big(2^{\gamma \,\CONS\, v_r}\big)
\Big]
\label{eq:Psi}
\end{equation}
wherein the multiindex $\lambda$ encapsulates log-wavelengths $\gamma \CONS v_r \in \mathbb{R}$ and particle spins $\theta \in \mathbb{T}$ and the infix operator $\CONS$ denotes list construction (``cons'') in the ML family of programming languages.
The conceptual jump from purely temporal scattering to time-frequency scattering eventually turned out to be fruitful, but difficult: because wavelengths $\gamma_m$  at one layer of the network (e.g.~pitch $\gamma_1$ or tempo $\gamma_2$) may take over the roles of spatial variables $v_r$ in a deeper network, keeping track of all cross-dependencies between variables appealed for a more systematic resort to recursion in our numerical applications.

And\'{e}n and myself studied the above definition in complementary ways.
He used the principle of stationary phase to confirm that time-frequency scattering characterizes the chirp rates of ripple stimuli, analogously to STRF in the primary auditory cortex.
He also designed a multiresolution analysis scheme for time-frequency scattering, in the fashion of Mallat's discrete wavelet transform algorithm and Simoncelli's steerable pyramid.
This scheme allowed to interpret the time-frequency scattering transform as the response of a deep convolutional neural network whose depth grows logarithmically with receptive field size.
On my part, I wrote down the production rules of the following context-sensitive grammar, so that the language of admissible paths in a time-frequency scattering network could be described exhaustively by a nondeterministic Turing machine with linearly bounded tape memory:

\begin{tabular}{r c l c r l l}
& $S$ & & $\rightarrow$ & & $t$ & \\
& $S$ & & $\rightarrow$ & & $t, (\gamma_1, X^*)?$ & \\
$\gamma_m$, & $X$ & & $\rightarrow$ & $\gamma_m$, & $\gamma_{m+1}, X^*$ & \\
$\gamma_m$, & $X$ & & $\rightarrow$ & $\gamma_m$, & $Y^n, \gamma_1\CONS\gamma_m, \theta_1\CONS\gamma_m, \gamma_{m+1}, X^n, X?$ & \qquad \qquad $(n\geq0)$\\
$\gamma_m$, & $Y,$ & $\gamma_k$& $\rightarrow$ & $\gamma_m$, & $\gamma_{k+1}\CONS\gamma_m,\theta_{k+1}\CONS\gamma_m,$ & $\gamma_k$
\\
$Y^n$, & $Y,$ & $\gamma_k \CONS \gamma_m$& $\rightarrow$ & $Y^n$, & $\gamma_{k+1}\CONS\gamma_m,\theta_{k+1}\CONS\gamma_m,$ & $\gamma_k \CONS \gamma_m.$ 
\end{tabular}
\\

Once the recursive grammar above was in place, I was able to reason at compile time on the computation graph of time-frequency scattering architectures, and cast Waldspurger's advances in phase retrieval from time scattering coefficients into a multivariable framework.
Upon advice from Joan Bruna, I opted for synthesizing sound by stochastic gradient descent: starting from a random initial guess --- usually, Brownian motion noise --- this procedure adds a corrective term to the signal at every iteration, so that its time-frequency scattering coefficients match those of a predefined textural target.
Incidentally, it is also by means of stochastic gradient descent that most of the algorithms that are known today, albeit somewhat improperly, as artificial intelligence, learn to perform tasks of computer vision, automatic speech recognition, and language translation.
Because time-frequency scattering networks, just like deep convolutional neural networks, consist of differentiable layers, the corrective term in stochastic gradient descent can be computed by a method of Lagrange multipliers, named backpropagation.
There is, however, one distinction between the two iterative procedures: whereas in deep learning, gradient backpropagation causes an infinitesimal update of synaptic weights in order to bring the predicted output closer to the ground truth, here, the synaptic weights are kept fixed, under the form of wavelet impulse response coefficients; but it is the raw waveform itself that gets updated towards a local minimum of the Euclidean error functional $\mathrm{E} = \Vert (\mathrm{E}_m)_m \Vert_2$, with
\begin{equation}
\mathrm{E}_m^2 =
\int\!\ldots\!\int_{\Lambda_1 \ldots \Lambda_m}
\left(
\int_{-\infty}^{+\infty} \mathbf{U}_m(\tau,\lambda_1 \ldots \lambda_m)^2\;\mathrm{d}\tau - \int_{-\infty}^{+\infty} \mathbf{U}_m^{\infty}(\tau,\lambda_1 \ldots \lambda_m)^2\;\mathrm{d}\tau
\right)
\;\mathrm{d}^m\lambda.
\label{eq:Em}
\end{equation}
Aside from this technical distinction, audio texture synthesis from scattering coefficients is quite comparable to the training of a deep neural network.
In both cases, the system produces uninformative outcomes at the start; and then, after being exposed to some real-world data, adjusts its own predictions by trial and error, until converging to a highly articulate statistical fit.

For Joakim And\'{e}n and myself, refactoring the source code of the software library for scattering transforms so that it could allow for multivariable architectures and gradient backpropagation, was a steady effort of almost two years, with many emotional ups and downs --- as often in scientific research.
By the end of 2015, we had a working implementation\footnote{github.com/lostanlen/scattering.m} and presented it at the IEEE conference on Machine Learning for Signal Processing (MLSP) in Boston \cite{anden2015mlsp}.
Our paper boiled down to three claims: first, time-frequency scattering is more mathematically interpretable than other auditory representations, be them engineered or learned; secondly, on some tasks for which the availability of annotated data is limited (e.g.~musical instrument recognition), it actually outperforms deep learning classifiers; and thirdly, it allows to reconstruct chirps in audio textures, such as bird vocalizations, with satisfying perceptual similarity to the target.
Yet, the section on signal re-synthesis was purely meant as an illustration of the capabilities and limitations of time-frequency scattering, as compared to other auditory representations.
Never in the research agenda of my PhD did I anticipate that time-frequency scattering could one day prove to be useful to contemporary music creation.

Florian Hecker wrote to me for the first time in the spring of 2016.
He had heard of time-frequency scattering through our mutual colleague Bob Sturm, and wanted to use it as a software for texture-related sound synthesis with  wavelets features.
When we first ran time-frequency scattering on his piece \emph{Modulator} (2014), I was pleased to find that it performed about as well in terms of perceptual similarity, while converging over 50 times faster.
Indeed, contrary to other STRF-inspired software, the time-frequency scattering library was using a multiresolution pyramid to spare unnecessary computations in the lower frequencies; moreover, the wavelet factorization in Equation \ref{eq:Psi} allowed to vectorize array operations and rely on fast Fourier transforms (FFT) to speed up convolutions.
These technical improvements, although leaving the gist of the algorithm essentially unchanged, noticeably streamlined the compositional workflow, by allowing rapid prototyping of ideas.
Because running one iteration of stochastic gradient descent now lasted about as long as the target sound clip, it became possible 
to listen to synthetic texture samples in real time, meanwhile time-frequency scattering was progressively converging towards a local optimum of Equation \ref{eq:Em}.

I opened this essay by depicting a schematic, and perhaps outdated, dichotomy between \emph{musique concr\`{e}te} and \emph{Elektronische Musik}.
I argued that both of these paradigms were following the same artistic research program --- that is, to liberate the Western canon from a thousand-year tradition of solmization that gives hegemonic power to the concept of musical note --- yet by clashing ways.
What \emph{musique concr\`{e}te} gained in terms of timbral sophistication, it lacked in terms of stylistic power.
Conversely, \emph{Elektronische Musik} achieved a maximal level of creative control, yet was restricted by a rudimentary collection of building blocks: pure tones.
This dilemma, as composer Jean-Claude Risset often said, was a direct consequence of the use of analog audio technologies.

Now in the age of digital information, the tradeoff between specificity and expressivity seems to have progressively softened, if not gone obsolete altogether.
In a piece such as Florian Hecker's \emph{FAVN} (2016), both traditions are kept alive in a perpetual \emph{jeu de miroirs} which dynamically alternates between the \emph{concr\`{e}te} paradigm (i.e.~to compute time-frequency scattering coefficients from the reconstructed waveform at iteration $n$) and the \emph{Elektronische} paradigm (i.e.~to synthesize a waveform at iteration $(n+1)$ from the numerical parameters obtained through gradient backpropagation at iteration $n$).
Then, once such a playful interaction is in place, the decision of printing out the values of time-frequency scattering coefficients, originating from an analysis of the three movements of FAVN,  figurates the \emph{ad infinitum} limit of both paradigms.

Between the analysis and re-synthesis steps, occurs a stage of abstraction: that of sorting all time-frequency scattering paths by the relative amount of energy that they carry.
Measuring energy in a given scattering path $\lambda$ is made possible by the Littlewood-Paley condition
\begin{equation}
    \forall \omega\!::\!v_r,\quad
    1 - \varepsilon \lesssim \boldsymbol{\widehat{\phi}}(\omega\!::\!v_r)^2 + \dfrac{1}{2} \sum_{\gamma::v_r} \big \vert \boldsymbol{\widehat{\psi}}_{\gamma::v_r} \big \vert (\omega\!::\!v_r)^2 \lesssim 1,
\end{equation}
which states that, for every variable $v_r$, the filterbank of wavelets $\boldsymbol{\psi}_{\gamma:: v_r}$ and its corresponding scaling function $\boldsymbol{\phi}$ unitarily cover the Fourier domain.
This double inequality implies that the amount of energy in a scattering representation is the same at every layer --- and, therefore, equal to the energy of the original waveform $\mathbf{U}_0(t)$.
Therefore, in the context of time-frequency scattering, and for any value of the path $\lambda = (\gamma_1, \gamma_2, \gamma_1\CONS\gamma_1)$, the ratio $\mathbf{U}_m (t, \lambda)$ is a dimensionless quantity between zero and one.
Multiplying this quantity by $10^6$ converts it into a number of parts per million (ppm).
This number is the leftmost column in the table.
The second column denotes acoustic frequency in Hertz (Hz), corresponding to the temporal log-frequency variable $\gamma_1$ in the first layer of the scattering network.
The third column denotes temporal modulation frequency, also known as \emph{rate} in Hertz (Hz), and corresponding to the temporal log-frequency variable $\gamma_2$ in the second layer.
It should be remarked that the acoustic frequency belongs to the audible range ($20$ Hz - $20$ kHz), but that the temporal modulation frequency can be as low as $1$ Hz, and as high as $1$ kHz under the condition $\gamma_1 < \gamma_2$.
Lastly, the fourth column denotes frequential modulation frequency, also known as \emph{scale} in cycles per octave (c/o), and corresponding to the variable $\gamma_1\CONS\gamma_1$ in the second layer.
With the mapping between time-frequency scattering paths $p=(\gamma_1,\gamma_2,\gamma_1\CONS\gamma_1)$ and averages energies in parts per million that is presented herein, there is enough information to replicate the auditory percepts of \emph{FAVN}, even in the absence of an waveform-domain record of the piece.

The numerical tables appearing in these pages epitomize one founding myth of computer music: that of a mental quest for ``the'' sound.
At the limit of technical feasibility, signal reconstruction is perfect and all phase incoherences have disappeared: the outcome is an exact, \emph{Elektronische} rendition of the original \emph{concr\`ete} material.
In other words, the procedure has gone full circle from \emph{Elektronische} to \emph{concr\`ete} and back, without alteration.
Nevertheless, owing to stochastic effects in the sampling of Brownian motion and the finiteness of computational resources, the sonified piece can only be a close approximate of its textual-numerical prototype.
In the to and fro of cognitive modeling and acoustic adjustment, the music of signals and the music of symbols chase each other like a cadenced farandole.
Quite paradoxically, the impact of mathematical quantization gradually becomes less noticeable as it becomes more accurate.

Here I do not mean to say, in what would be a paraphrase of Leibniz, that ``Music is a hidden arithmetic exercise of the soul, which does not know that it is counting''.
I do not, either, mean that the numeric tables that are printed herein could aspire to be a proxy for the auditory experience: on the contrary, I firmly believe that music is meant to be heard, and that no other medium can replace it, or even refer to it in any formal ``word-object'' correspondence system.
Thirdly, I do not think of music as a language in the same sense as our other forms of communication, be them spoken, written, or signed; and therefore certainly not of this publication as an ersatz of post-serialist musical score.
Rather, and despite the utter ineffability of music, it is possible to shed light upon our shared faculty of recursion, supplemented by perceptual quantization and tabular organization; of which musical notation is a mere by-product.

Far from any neo-numerological considerations, what is, in my mind, the intimate \emph{raison d'\^etre} of this publication, is that it helps us listeners understand two compositional prospects, and wraps them into one: the will to expand the scope of the potentially audible, by seeking for more and more complexity in the parametrization of sound synthesis; and the desire to delve deeper into what has been heard, by shifting the auditory focus onto previously unnoticed details.
Music is, therefore, a two-fold ritual of anticipation.
Like the composer, it is in the liminality of finite speeds that the faun shall dwell and thrive.

\section*{Acknowledgment}

This work is supported by the ERC InvariantClass 320959. The author wishes to thank Th\'eis Bazin, Graham Dove, Jan Ferreira, and Lorenzo Senni for helpful discussions.

\bibliography{references}
\bibliographystyle{ieeetr}

\end{document}